# Two Pressure-induced Superconducting Transitions in SnBi$_2$Se$_4$ Explored by Data-driven Materials Search: New Approach to Develop Novel Functional Materials Including Thermoelectric and Superconducting Materials

Ryo Matsumoto[1,4]*, Zhufeng Hou[2], Hiroshi Hara[1,4], Shintaro Adachi[1], Hiroyuki Takeya[1], Kiyoyuki Terakura[3] and Yoshihiko Takano[1,4]

[1]International Center for Materials Nanoarchitectonics (MANA),

National Institute for Materials Science, 1-2-1 Sengen, Tsukuba, Ibaraki 305-0047, Japan

[2]Research and Services Division of Materials Data and Integrated System (MaDIS),

National Institute for Materials Science, 1-2-1 Sengen, Tsukuba, Ibaraki 305-0047, Japan

[3]Center for Materials research by Information Integration (CMI[2]),

National Institute for Materials Science, 1-2-1 Sengen, Tsukuba, Ibaraki 305-0047, Japan

[4]University of Tsukuba, 1-1-1 Tennodai, Tsukuba, Ibaraki 305-8577, Japan

E-mail: MATSUMOTO.Ryo@nims.go.jp

Candidates for new thermoelectric and superconducting materials, which have narrow band gap and flat bands near band edges, were searched by the high-throughput first-principles calculation from an inorganic materials database. The synthesized SnBi$_2$Se$_4$ among the target compounds showed a narrow band gap of 354 meV, and a thermal conductivity of ~1 W・K$^{-1}$m$^{-1}$ at ambient pressure. The sample SnBi$_2$Se$_4$ showed a metal-insulator transition at 11.1 GPa, as predicted by a theoretical estimation. Furthermore, the two pressure-induced superconducting transitions were discovered at under 20.2 GPa and 47.3 GPa. The data-driven search is a promising approach to discover new functional materials.

Data-driven material science (materials informatics [1], materials genome initiative [2], chemometrics [3], and so on) recently brings remarkable results [4-10]. For example, the discovery of new cathode materials to extend the lifetime of lithium ion battery can be accelerated by using high-throughput screening approach [11]. On the other hand, searching for new functional materials of thermoelectric materials and/or superconductors has been mostly still conducted through a carpet-bombing type experiment depending on experiences and inspirations of researcher.

In this study, we exhaustively searched the candidates for new thermoelectric and superconducting materials by first-principles calculation using a guideline which is characterized by specific band structures of "flat band" near the Fermi level, such as multivalley [12], pudding mold [13], and topological-type [14] structures. If such kinds of flat band approach Fermi level, thermoelectric properties of electrical conductivity and Seebeck coefficient would be enhanced [13,15]. If the flat band crosses the Fermi level, superconductivity would be realized due to high density of states (DOS) [16-18].

Single crystal samples of the compound chosen by the above-mentioned screening were successfully synthesized. The transport properties of the obtained sample were evaluated under ambient and high pressure. Here we report the discoveries of not only the pressure-induced metal-insulator transition of the sample but also superconducting phases under further pressure. This work successfully demonstrates an exploration of new functional materials by data-driven search.

To search new thermoelectric and superconducting materials, we conducted the high-throughput first-principles calculations on about 1570 candidates of ternary compounds listed from the inorganic material database named AtomWork [19]. The selection of these candidate compounds was guided by the following restriction: abundant and nontoxic or less toxic constituent elements, and the number of atoms less than 16 per primitive unit cell. The first-principles calculations were carried out using the projector-augmented wave (PAW) method [20] within the generalized gradient approximation (GGA), in which the Perdew-Burke-Ernzerhof (PBE) [21] exchange-correlation functional was used. For every candidate compound, we first carried out four steps of first-principles calculations including structure relaxation, self-consistent field (SCF) calculations, band structure, and DOS of its crystal structure. From the calculated band structures and DOSs, we narrowed down the number of candidates by selecting the compounds that shall have real or pseudo energy gap. Additionally, a certain size (e.g., ≤ 0. 6 eV in the GGA-PBE calculations) in the band gap, high DOSs near the Fermi level or the valence and conduction band edges, and the flat band were considered into the selection criteria. By this screening, the number of candidate compounds was reduced to 45. Then we carried out the aforementioned four steps of first-principles calculations for these 45 compounds under a high pressure of 10 GPa. By checking whether the band gap is decreased (or even the metallic behavior appears) or not, we screened out 27 promising compounds for our further studies to uncover their functionality. Through the above screening procedures, $SnBi_2Se_4$ was chosen as a candidate for new thermoelectric and superconducting materials. The details of our screening scheme in the high-throughput first-principles calculations and the complete results will be given elsewhere.

Figure 1 shows (a) crystal structure of $SnBi_2Se_4$ (trigonal *R-3m* structure, lattice constants of $a = b = 4.188$ Å and $c = 39.46$ Å [22] ), drown by VESTA [23], (b) the band structure and (c) total DOS of $SnBi_2Se_4$ by the GGA-PBE calculations with the spin-orbital coupling. We can see that $SnBi_2Se_4$ has a narrow band gap of ~270 meV and that both the top valence band along several k-points paths and the bottom conduction band along Γ-Z are the quite flat. The valence bands especially show high total DOS near the Fermi level. The top valence band is contributed mainly by the Se $4p$ states and the Sn $5s$ states, while the bottom conduction band is contributed mainly by the Bi $6s$ states. Such a band structure satisfies the screening condition. The band structures under high pressure suggest that the band edges approach the Fermi level around 5-10 GPa, indicating a pressure-induced metal-insulator transition.

Single crystals of $SnBi_2Se_4$ were grown by a melt and slow-cooling method. Starting materials of Bi grains (99.9%), Sn grains (99.99%) and Se chips (99.999%) were put into an evacuated quartz tube in the stoichiometric composition of $SnBi_2Se_4$. The ampoule was heated at 300ºC for 8 hours, subsequently at 550ºC for 10 hours. The obtained powders were ground and loaded into an evacuated quartz tube. The sample was heated again at 700ºC for 10 hours, and slowly cooled to 620ºC for 8 hours followed by furnace cooling.

The crystal structure of the obtained sample was investigated by a powder X-ray diffraction (XRD) using a Mini Flex 600 (Rigaku). The chemical composition of the sample was evaluated by single-crystal structural analysis using a XtaLAB mini (Rigaku) and an energy dispersive X-ray spectrometry (EDX) analysis using a JSM-6010LA (JEOL). Figure 2 shows the powder XRD pattern of the pulverized sample. Most of the observed peaks were well indexed to trigonal *R-3m* structure with lattice constants of $a = b = 4.26$ Å and $c = 39.17$ Å, except for a few weak peaks of impurity phases. The single crystal structural analysis revealed that the actual composition is $SnBi_{1.72}Se_{3.54}$ calculated from the site occupancies.

The EDX analysis showed the composition of SnBi$_{1.64}$Se$_{3.53}$, which is in accordance with the single crystal structural analysis.

The thermoelectric properties of the obtained SnBi$_2$Se$_4$ were investigated. Figure 3 shows the temperature dependences of (a) electrical resistivity, (b) Seebeck coefficient, (c) carrier concentration, and (d) thermal conductivity under ambient pressure, measured by use of physical property measurement system (Quantum Design: PPMS) with thermal transport option. According to the slope of Arrhenius plot for the resistivity, the sample was determined to have a narrow band gap of 354 meV near room temperature, which is consistent with the GGA-PBE calculations. The negative Seebeck coefficient and negative slope of the Hall voltage indicate n-type nature of the sample, which may be caused by the vacancy defects. The thermal conductivity of ~1 W·K$^{-1}$m$^{-1}$ was remarkably low even near room temperature. On the other hand, the small power factor of 9.1 μWm$^{-1}$K$^{-2}$ and figure of merit ZT of 0.0038 near room temperature were observed as shown in Fig.3 (e-f), caused by its high resistivity. It is possible to improve these thermoelectric properties by decrease of resistivity using carrier doping [24], or high pressure application [25-27]. The calculations of power factor based on the band structures under ambient and high pressure using a BoltzTrap code [28] suggests that the power factor could be enhanced by applying pressure around the metal-insulator transition, due to a decrease of resistivity. We therefore focused on the resistivity of the SnBi$_2$Se$_4$ under high pressure.

To explore the pressure-induced metal-insulator transition and superconductivity in SnBi$_2$Se$_4$, we performed electrical resistivity measurement under high pressure using an originally designed diamond anvil cell with boron-doped diamond electrodes [29-32]. The SnBi$_2$Se$_4$ single crystal was placed at the center of bottom anvil where the boron-doped diamond electrodes were fabricated. Except for the sample space and electrical terminal, the surface of the bottom anvil was covered by undoped diamond insulating layer. The detail of the cell configuration is described in the literatures [31,32]. The cubic boron nitride powders with ruby manometer were used as a pressure-transmitting medium. The applied pressure values were estimated by the fluorescence from ruby powders [33] and the Raman spectrum from the culet of top diamond anvil [34] by an inVia Raman Microscope (RENISHAW).

Figure 4(a) shows the temperature dependence of resistance for the SnBi$_2$Se$_4$ under the various pressures from 1.9 GPa to 16.9 GPa. The band gaps estimated from Arrhenius plot near room temperature were labeled at each pressure. The resistance and the band gap decreased with the increase of the applied pressure. A metallic conductivity of the sample appeared at 11.1 GPa, indicating a pressure-induced metal-insulator transition. This behavior is qualitatively consistent with the theoretical calculation. The high thermoelectric performance of this compound would be expected at pressure around the metal-insulator transition, since the flat bands with high DOS approach the Fermi level.

Figure 4(b) shows the temperature dependence of resistance from 20.2 GPa to 42.9 GPa. A pressure-induced superconductivity with clear zero resistance was observed above 20.2 GPa. The maximum onset transition temperature ($T_c^{onset}$) and zero-resistance temperature ($T_c^{zero}$) were 2.5 K and 2.1 K under 42.9 GPa, respectively. The critical current $I_c$ and upper critical field $H_{c2}^{//ab}(0)$ were 35 μA and 2.1 T under 42.9 GPa, respectively, as shown in the Fig. 5. The $H_{c2}^{//ab}(0)$ value was estimated from the Werthamer-Helfand-Hohenberg (WHH) approximation [35] for the Type II superconductor in a dirty limit. The $T_c$, $I_c$, and $H_{c2}^{//ab}(0)$ values in this superconducting phase were almost independent to pressure.

Surprisingly, the $T_c$ of SnBi$_2$Se$_4$ jumped up with the application of further pressure which implies the discovery of a second superconducting phase. The temperature dependence of resistance from 42.9 GPa to 71.6 GPa were summarized in the Fig. 4(c). The $T_c^{onset}$ was suddenly enhanced from 2.5 K under 42.9 GPa to 5.9 K under 63.2 GPa. In the higher $T_c$ phase, the maximum $I_c$ and $H_{c2}^{//ab}(0)$ were enhanced up to 525 μA and 3.2 T under 63.2GPa, compared to those of thelower $T_c$ phase.

Figure 6 shows the pressure phase diagram of SnBi$_2$Se$_4$ single crystal. The resistance of the sample is dramatically decreased by applying pressure, and metal-insulator transition occurred at 11.1 GPa. In this region, high performance thermoelectric property can be expected in this sample. After that, the first superconducting phase was newly discovered under 20.2 GPa. The $T_c$ was constant up to 42.9 GPa. In the pressure region higher than 47.3 GPa, we observed a higher $T_c$ phase. This superconductivity survived up to at least 71.6 GPa.

In conclusion, a new thermoelectric and superconducting material was successfully discovered by data-driven materials search using a combination of first-principles calculation and high pressure technique. The high thermoelectric property of the chosen material SnBi$_2$Se$_4$ was simulated under high pressure around metal-insulator transition. The synthesized SnBi$_2$Se$_4$ single crystal showed two pressure-induced superconducting transitions with a maximum $T_c$ of 5.9 K under 63.2 GPa. The present work will serve as a case study of the important first-step for the next generation data-driven material science.

**Acknowledgments**

This work was partly supported by JST CREST, Japan, JST-Mirai Program Grant Number JPMJMI17A2, Japan, JSPS KAKENHI Grant Number JP17J05926, and the "Materials research by Information Integration" Initiative (MI$^2$I) project of the Support Program for Starting Up Innovation Hub from JST. A part of the fabrication process of diamond electrodes was supported by NIMS Nanofabrication Platform in Nanotechnology Platform Project sponsored by the Ministry of Education, Culture, Sports, Science and Technology (MEXT), Japan. The part of the high pressure experiments were supported by the Visiting Researcher's Program of Geodynamics Research Center, Ehime University. The computation in this study was performed on Numerical Materials Simulator at NIMS.

**References**

1) John R. Rodgers and David Cebon, MRS Bull. **31**, 975 (2006).
2) A. Jain, S. P. Ong, G. Hautier, W. Chen, W. D. Richards, S. Dacek, S. Cholia, D. Gunter, D. Skinner, G. Ceder, and K. A. Persson, APL Mater. **1**, 011002 (2013).
3) B. R. Kowalski, Trends Anal. Chem. **1**, 71 (1981).
4) Y. Hinuma, T. Hatakeyama, Y. Kumagai, L. A. Burton, H. Sato, Y. Muraba, S. Iimura, H. Hiramatsu, I. Tanaka, H. Hosono, and F. Oba, Nat. Commun. **7**, 11962 (2016).
5) R. Jalem, M. Nakayama, and T. Kasuga, J. Matter. Chem. A **2**, 720 (2014).
6) R. Jalem, K. Kanamori, I. Takeuchi, M. Nakayama, H. Yamasaki, and T. Saito, Sci. Rep. **8**, 5845 (2018).
7) S. Kiyohara, H. Oda, T. Miyata, and T. Mizoguchi, Sci. Adv. **2**, e1600746 (2016).
8) A. Seko, A. Togo, H. Hayashi, K. Tsuda, L. Chaput, and I. Tanaka, Phys. Rev. Lett. **115**, 205901 (2015).
9) T. Inoshita, S. Jeong, N. Hamada, and H. Hosono, Phys. Rev. X **4**, 031023 (2014).
10) S. P. Ong, Y. Mo, W. D. Richards, L. Miara, H. S. Leeb, and G. Ceder, Energy Environ. Sci. **6**, 148 (2013).
11) M. Nishijima, T. Ootani, Y. Kamimura, T. Sueki, S. Esaki, S. Murai, K. Fujita, K. Tanaka, K. Ohira, Y. Koyama, and I. Tanaka, Nat. Commun. **5**, 4553 (2014).
12) D. M. Rowe, CRC handbook of thermoelectrics, (CRC Press, 2010).
13) K. Kuroki and R. Arita, J. Phys. Soc. Jpn. **76**, 083707 (2007).
14) N. B. Kopnin, T. T. Heikkilä, and G. E. Volovik, Phys. Rev. B **83**, 220503 (2011).
15) K. Mori, H. Usui, H. Sakakibara, and K. Kuroki, J. Phys. Soc. Jpn, **83**, 023706 (2014).

16) K. S. Fries and S. Steinberg, Chem. Mater. **30**, 2251 (2018).

17) W. Sano, T. Koretsune, T. Tadano, R. Akashi, and R. Arita, Phys. Rev. B **93**, 094525 (2016).

18) Y. Ge, F. Zhang, and Y. Yao, Phys. Rev. B **93**, 224513 (2016).

19) Y. Xu, M. Yamazaki, and P. Villars, Jpn. J. Appl. Phys. **50**, 11RH02 (2011).

20) G. Kresse and D. Joubert, Phys. Rev. B **59**, 1758(1999).

21) J. P. Perdew, K. Burke and M. Ernzerhof, Phys. Rev. Lett. **77**, 3865 (1996).

22) A. A. Sher, I. N. Odin, A. V. Novoselova, Inorg. Mater. **14**, 993 (1978).

23) K. Momma and F. Izumi, J. Appl. Crystallogr. **44**, 1272 (2011).

24) A. T. Duong, V. Q. Nguyen, G. Duvjir, V. T. Duong, S. Swon, J. Y. Song, J. K. Lee, J. E. Lee, S. Park, T. Min, J. Lee, J. Kim, and S. Cho, Nat. Commun. **7**, 13713 (2016).

25) R. G. Zhong, J. X. Peng, Z. P. Wen, Z. C. Yi, M. H. An, and W. X. Cheng, Chin. Phys. Lett. **22**, 236, (2005).

26) S. V. Ovsyannikov, V. V. Shchennikov, G. V. Vorontsov, A. Y. Manakov, A. Y. Likhacheva, and V. A. Kulbachinskii, J. Appl. Phys. **104**, 053713 (2008).

27) R. S. Kumar, M. Balasubramanian, M. Jacobsen, A. Bommannavar, M. Kanatzidis, S. Yoneda, and A. L. Cornelius, J. Electro. Matter. **39**, 1828 (2010).

28) G. K. H. Madsena and D. J. Singh, Comput. Phys. Commun. **175**, 67 (2006).

29) R. Matsumoto, Y. Sasama, M. Fujioka, T. Irifune, M. Tanaka, T. Yamaguchi, H. Takeya, and Y. Takano, Rev. Sci. Instrum. **87**, 076103 (2016).

30) R. Matsumoto, T. Irifune, M. Tanaka, H. Takeya, and Y. Takano, Jpn. J. Appl. Phys. **56**, 05FC01 (2017).

31) R. Matsumoto, A. Yamashita, H. Hara, T. Irifune, S. Adachi, H. Takeya, and Y. Takano, Appl. Phys. Express **11**, 053101 (2018).

32) R. Matsumoto, H. Hara, H. Tanaka, K. Nakamura, N. Kataoka, S. Yamamoto, T. Irifune, A. Yamashita, S. Adachi, H. Takeya, and Y. Takano, (In preparation).

33) G. J. Piermarini, S. Block, J. D. Barnett, and R. A. Forman, J. Appl. Phys. **46**, 2774 (1975).

34) Y. Akahama and H. Kawamura, J. Appl. Phys. **96**, 3748 (2004).

   N. R. Werthamer, E. Helfand, and P. C. Hohenberg, Phys. Rev. **147**, 295 (1966).

# Figure Captions

**Fig. 1.** (a) Crystal structure, (b) band structure and (c) density of states (DOS) of SnBi$_2$Se$_4$ obtained by the GGA-PBE calculations with the spin-orbital coupling.

**Fig. 2.** Powder XRD pattern of pulverized SnBi$_2$Se$_4$ single crystal.

**Fig. 3.** Temperature dependence of thermoelectric properties in SnBi$_2$Se$_4$. (a) resistivity (inset is Arrhenius plot), (b) Seebeck coefficient, (c) carrier concentration (inset is a magnetic field dependence of Hall voltage at room temperature), (d) thermal conductivity, (e) power factor, and (f) figure of merit ZT.

**Fig. 4.** Temperature dependence of resistance in SnBi$_2$Se$_4$ under various pressures, (a) 1.9 - 16.9 GPa, (b) 20.2 - 42.9 GPa, and (c) 42.9 - 71.6 GPa. Insets show enlargement around $T_c$.

**Fig. 5.** Highest superconducting properties in lower $T_c$ phase and higher $T_c$ phase. (a) current-voltage curve at 42.9 GPa and 63.2 GPa, (b) $H_{c2}^{//ab}$ curve at 42.9 GPa and 63.2 GPa.

**Fig. 6.** Pressure phase diagram in SnBi$_2$Se$_4$.

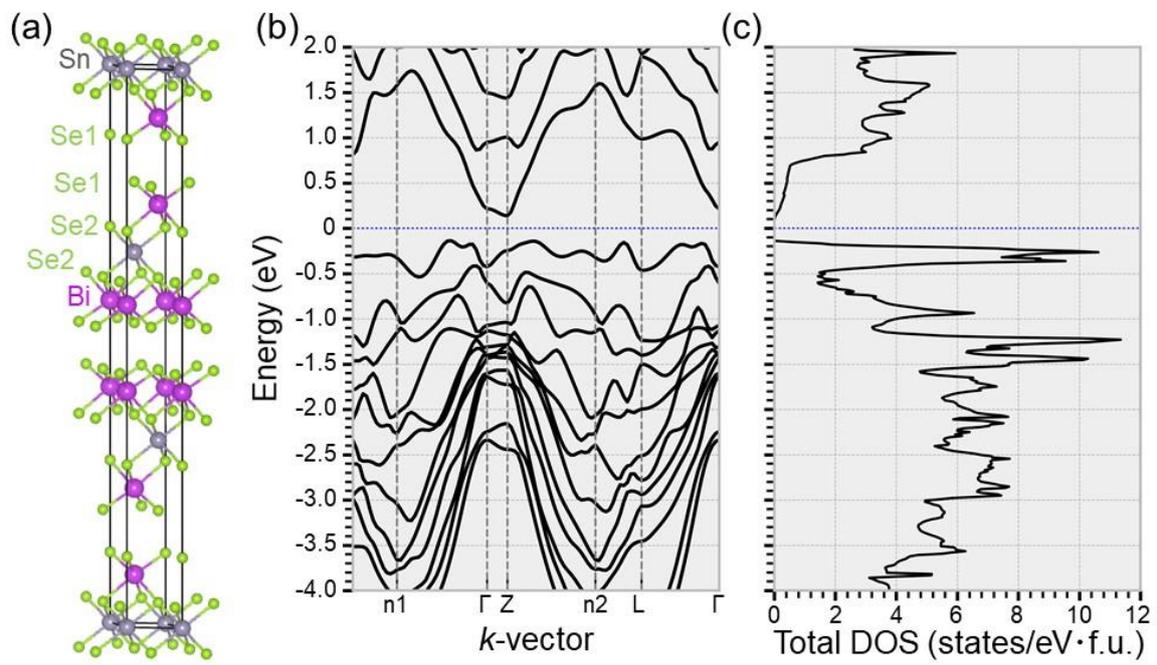

Fig.1.

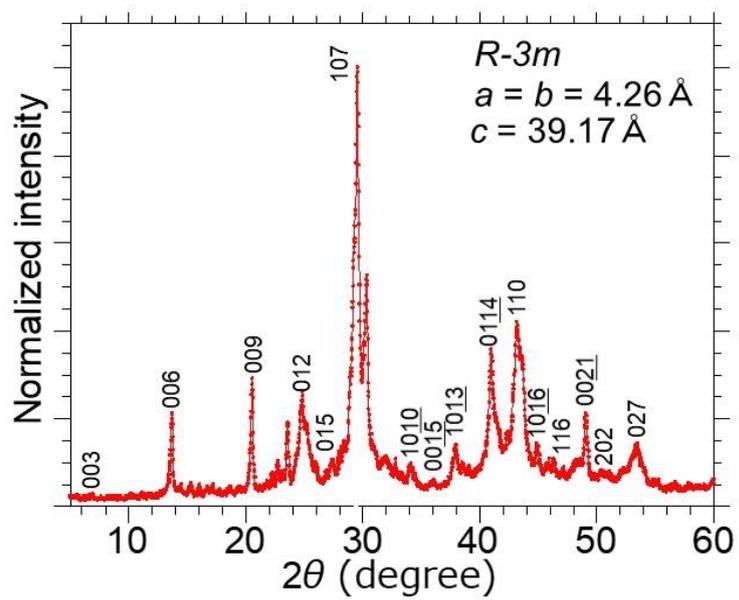

Fig.2.

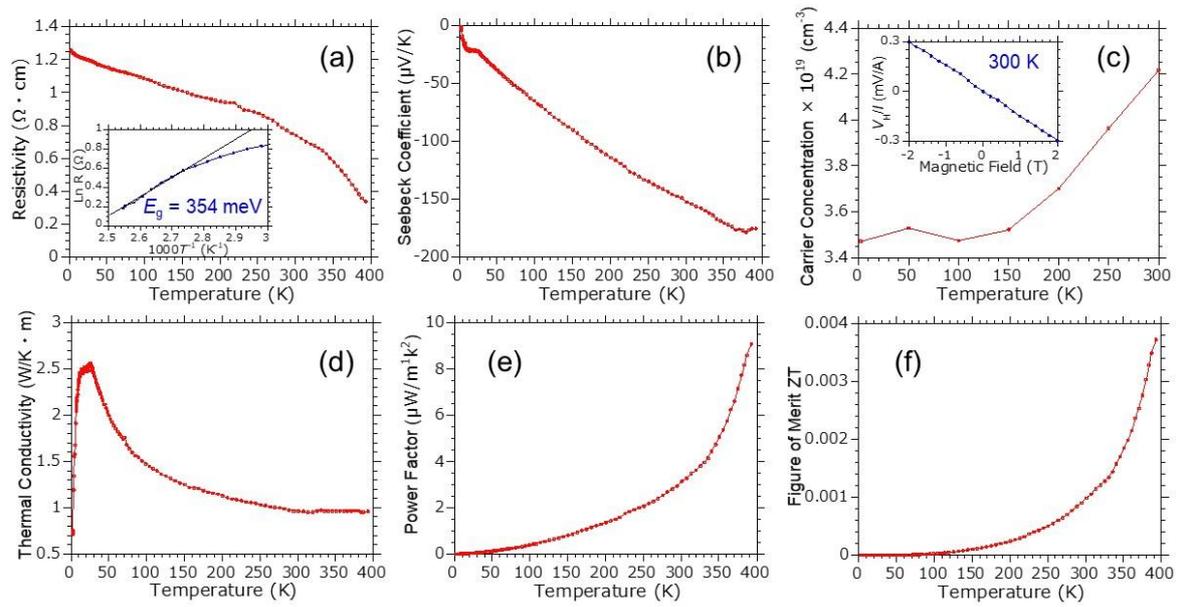

Fig.3.

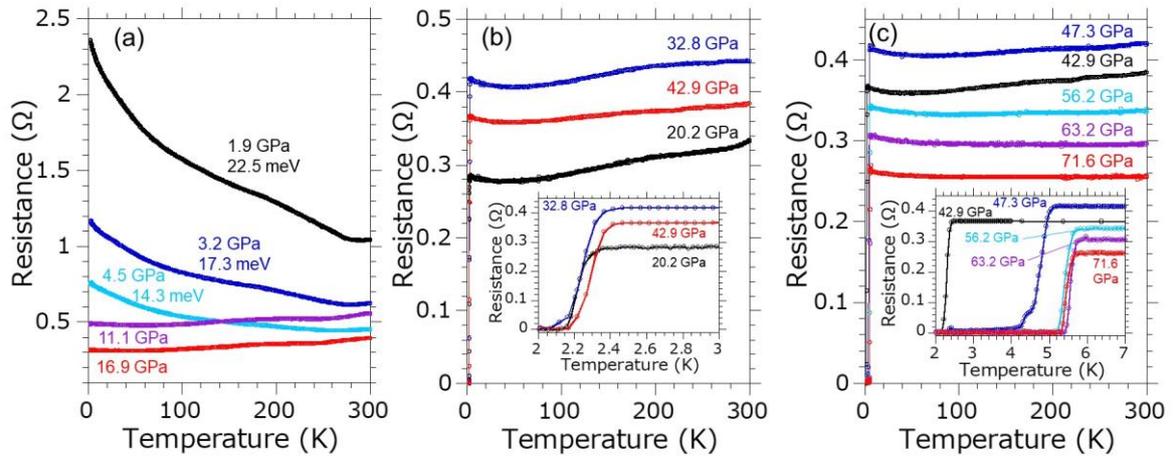

Fig.4.

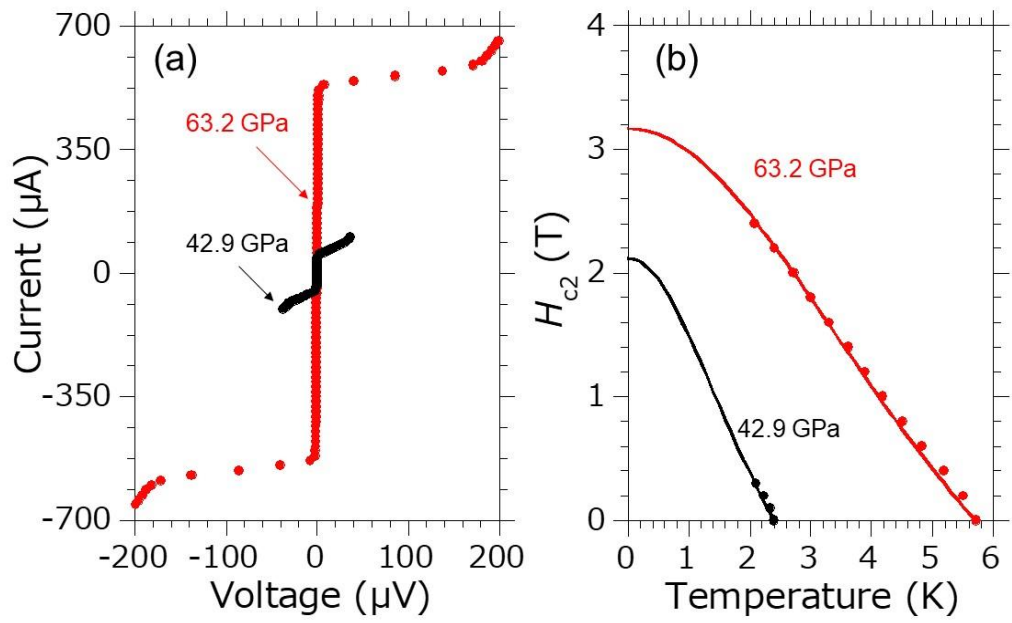

Fig.5.

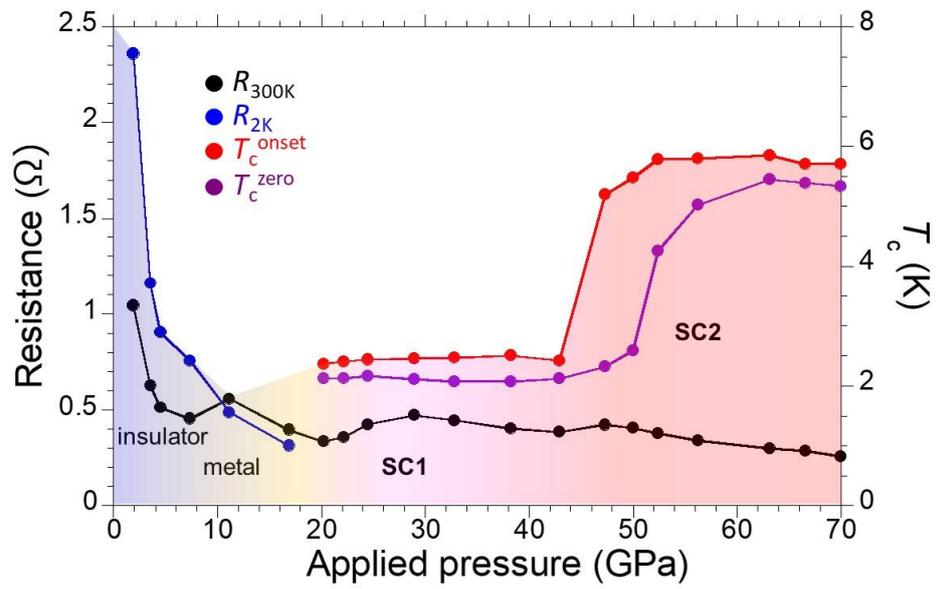

Fig.6.